\documentclass[superscriptaddress,amsmath,amssymb,prd,nofootinbib,twocolumn]{revtex4-2}

\usepackage{mathrsfs}
\usepackage[x11names]{xcolor}
\usepackage{graphicx}
\usepackage{dcolumn}
\usepackage{bm}
\usepackage[pdftex]{hyperref}
\usepackage{soul}
\usepackage{float}
\usepackage{orcidlink}

\hypersetup{pdftitle={Galactic Diffuse Neutrino Emission from Sources beyond the Discovery Horizon},
pdfsubject={Galactic Diffuse Neutrino Emission},
pdfauthor={Ambrosone, Groth, Peretti \& Ahlers},
pdfstartview={FitH},
colorlinks=true,
bookmarksopen=false,
bookmarksnumbered=false,
bookmarksopenlevel=0,
linkcolor=Blue1!60!black,
citecolor=Green1!50!black,
urlcolor=Blue1!70!black
}

\begin{document}

\title{Galactic Diffuse Neutrino Emission from Sources beyond the Discovery Horizon}

\author{Antonio Ambrosone\,\orcidlink{0000-0002-9942-1029}}
\email{antonio.ambrosone@unina.it}
\affiliation{Dipartimento di Fisica ``Ettore Pancini'', Universit\`{a} degli studi di Napoli ``Federico II'', Complesso Univ.~Monte S.~Angelo, I-80126 Napoli, Italy}
\affiliation{INFN - Sezione di Napoli, Complesso Univ.~Monte S. Angelo, I-80126 Napoli, Italy}
\author{Kathrine Mørch Groth\,\orcidlink{0000-0002-1581-9049}}%
 \email{kathrine.groth@nbi.ku.dk}
\affiliation{Niels Bohr International Academy, Niels Bohr Institute, University of Copenhagen,\\ Blegdamsvej 17, DK-2100 Copenhagen, Denmark}
\author{Enrico Peretti\,\orcidlink{0000-0003-0543-0467}}
\affiliation{Niels Bohr International Academy, Niels Bohr Institute, University of Copenhagen,\\ Blegdamsvej 17, DK-2100 Copenhagen, Denmark}%

\author{Markus Ahlers\,\orcidlink{0000-0003-0709-5631}}
\affiliation{Niels Bohr International Academy, Niels Bohr Institute, University of Copenhagen,\\ Blegdamsvej 17, DK-2100 Copenhagen, Denmark}

\date{February 6, 2024}

\begin{abstract}
The IceCube Neutrino Observatory has recently reported strong evidence for neutrino emission from the Galactic plane. The signal is consistent with model predictions of diffuse emission from cosmic ray propagation in the interstellar medium. However, due to IceCube's limited potential of identifying individual neutrino sources, it is also feasible that unresolved Galactic sources could contribute to the signal. We investigate the contribution of this quasi-diffuse emission and show that the observed Galactic diffuse flux at 100~TeV could be dominated by hard emission of unresolved sources. Particularly interesting candidate sources are young massive stellar clusters that have been considered as cosmic-ray PeVatrons. We examine whether this hypothesis can be tested by the upcoming KM3NeT detector or the planned future facility IceCube-Gen2 with about five times the sensitivity of IceCube.
\end{abstract}


\maketitle

\section{Introduction}\label{sec1}

Cosmic rays (CRs) with energies up to a few PeV are expected to originate in Galactic sources; see {\it e.g.}~Refs.~\cite{Blasi:2013rva,Amato:2017dbs,Gabici:2019jvz} for recent reviews. This hypothesis can be indirectly tested by observing the emission of $\gamma$-rays and neutrinos associated with the collisions of CRs with gas in the vicinity of their sources or while they propagate through the interstellar medium (ISM). Indeed, $\gamma$-ray observatories have detected a plethora of Galactic $\gamma$-ray sources \cite{Fermi-LAT:2017mzh,HESS:2018pbp,HAWC:2019tcx,LHAASO:2021gok} as well as extended diffuse emission \cite{Fermi-LAT:2012edv,Fermi-LAT:2015sau,TibetASgamma:2021tpz,Zhao:2021dqj,lhaaso:2023gne}, which can be attributed, in part, to the presence of CRs. However, the interpretation of these observations requires a careful modeling of absorption processes as well as the inclusion of $\gamma$-rays from synchrotron emission, bremsstrahlung, or inverse-Compton scattering of high-energy electrons.

\begin{figure}[b!]\centering
\includegraphics[width=\linewidth]{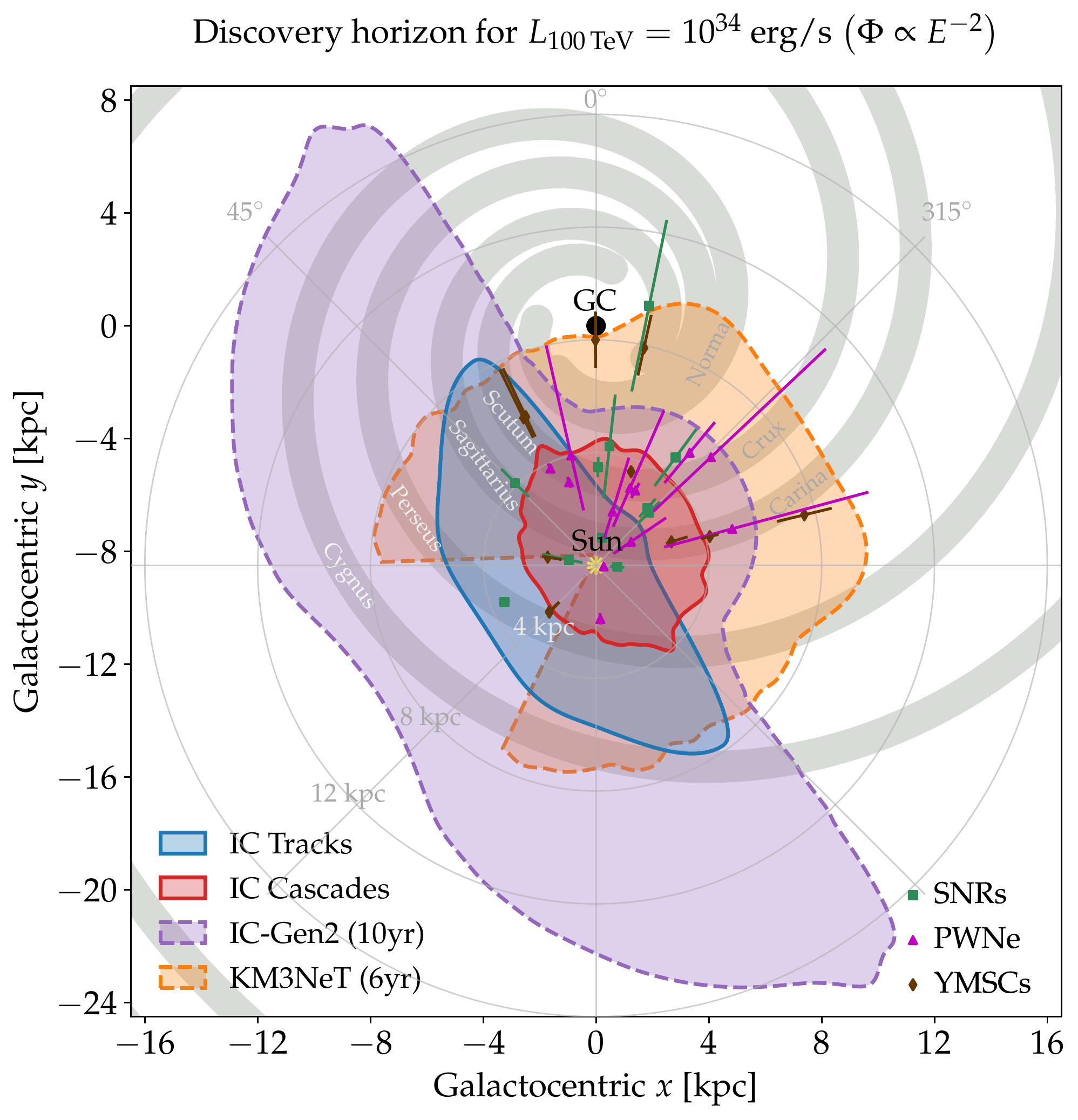}\\[0cm]
\caption[]{  IceCube's detection horizon for Galactic neutrino sources with an $E^{-2}$ emission spectrum (``IC Tracks''~\cite{IceCube:2019cia} and ``IC Cascades''~\cite{IceCube2023galactic}) and the expected reach of KM3NeT~\cite{KM3NeT:2018wnd} and the proposed IceCube-Gen2 facility~\cite{IceCube-Gen2:2020qha,Gen2_TDR} assuming a monochromatic neutrino luminosity $L_{100\,{\rm TeV}} = 10^{34}~{\rm erg/s}$. We indicate the location of Galactic arms~\cite{Steiman-Cameron:2010iuq} and nearby candidate neutrino sources. See main text for details.}\label{fig:DP_MW}
\end{figure}

In a recent study~\cite{IceCube2023galactic}, the IceCube experiment reported the first observation of high-energy neutrino emission from the Galactic plane (GP) with a significance of $4.5\sigma$. The result is based on a fit of neutrino emission templates derived from models of CR propagation and interaction in the Milky Way~\cite{Fermi-LAT:2012edv,Gaggero:2015xza}. The best-fit normalization of the angular-integrated per-flavor neutrino flux is at the level of $E_\nu^2\Phi \simeq 2\cdot10^{-8}~{\rm GeV}{\rm cm}^{-2}{\rm s}^{-1}$ at a neutrino energy $E_\nu=100$~TeV and marginally consistent with model predictions; see {\it e.g.}~Ref.~\cite{Gaggero:2015xza}. 

The IceCube analysis~\cite{IceCube2023galactic} is based on a selection of cascade events, {\it i.e.}~events with compact Cherenkov-light features following from a cascade of secondary short-ranged particles. Since these events have a relatively high angular uncertainty of typically $7^\circ$, the analysis has a limited ability to resolve degree-scale emission from individual neutrino sources. In other words, the diffuse Galactic neutrino flux measured by IceCube could in fact be a combination of a truly diffuse emission from hadronic interactions of CRs in the ISM, and the combined emission from unresolved Galactic sources which would appear as a quasi-diffuse flux.

In the following, we investigate the contribution of unresolved Galactic neutrino sources to the Galactic diffuse flux. Previous studies have already investigated the Galactic neutrino emission from both unresolved Galactic sources as well as from CR propagation~\cite{Ahlers:2013xia,Mandelartz:2013evw,Anchordoqui:2014rca,Ahlers:2015moa,Fang:2021ylv,Sudoh:2022sdk,Schwefer:2022zly,Sudoh:2023qrz}. Analogous to the case of Galactic TeV $\gamma$-ray sources~\cite{Cataldo:2020qla,Steppa:2020qwe,Vecchiotti:2021yzk}, the relative contribution of unresolved sources to the Galactic diffuse emission is expected to increase with energy due to the relatively soft emission from CRs in the interstellar medium~\cite{Stecker:1978ah,Domokos:1991tt,Berezinsky:1992wr,Ingelman:1996md,Evoli:2007iy,Ahlers:2013xia,Neronov:2013lza,Joshi:2013aua,Kachelriess:2014oma,Ahlers:2015moa,Gaggero:2015xza,Fang:2021ylv,Schwefer:2022zly,DelaTorreLuque:2022ats}. 
 
 We present here a novel model-independent formalism that parametrizes the (quasi-)diffuse Galactic emission in terms of the effective source surface density and neutrino luminosity, motivated by previous works on extragalactic neutrino populations, {\it e.g.}~Refs.~\cite{Ahlers:2014ioa,Murase:2016gly,Ackermann:2019ows}. We show that IceCube's limited discovery potential for individual neutrino sources -- in particular for extended Galactic sources -- permits a strong contribution of unresolved sources to the Galactic diffuse flux at 100~TeV. The upcoming KM3NeT ARCA~\cite{KM3Net:2016zxf,KM3NeT:2018wnd} and the planned IceCube-Gen2~\cite{IceCube-Gen2:2020qha,Gen2_TDR} have the potential to probe this hypothesis. 

\section{Quasi-Diffuse Neutrino Emission}\label{sec2}

The expected quasi-diffuse (QD) flux of unresolved Galactic neutrino sources depends on their spatial distribution as well as the variability of their emission spectra. For simplicity, we will assume in the following that neutrino sources are standard candles with a fixed luminosity following a power-law per-flavor neutrino spectrum $Q_\nu \propto E_\nu^{-\gamma}$ (units of ${\rm GeV}^{-1}\,{\rm s}^{-1}$) in the TeV--PeV energy range. The combined QD flux $\phi_{\rm QD}$ (units of ${\rm GeV}^{-1}\,{\rm cm}^{-2}\,{\rm s}^{-1}$) from sources along the line of sight ${\bf n}(\Omega)$ is then given via the integral:
\begin{equation}\label{eq:quasi_diffuse}
\phi_{\rm QD}(E_\nu,\Omega) = \frac{Q_{\nu}(E_\nu)}{4\pi}\int_0^\infty{d} D \rho({\bf r}_\odot + D{\bf n}(\Omega))\,,
\end{equation}
where $\rho({\bf r})$ is the source density, $D$ is distance from the Solar System to the source, $\Omega$ is the solid angle and ${\bf r}_\odot$ represents the location of the Solar System in the GP with distance $r_\odot \simeq 8.5$~kpc to the Galactic Center (GC). Figure~\ref{fig:DP_MW} shows our relative location with respect to the GC, Galactic arms and nearby candidate neutrino sources, including pulsar wind nebulae (PWNe), supernova remnants (SNRs), and young massive stellar clusters (YMSCs).

We consider in the following source distributions that align with the GP and are therefore effectively two-dimensional over large distances. As our fiducial model, we will assume azimuthally symmetric distributions that can be parametrized as $\rho({\bf r}) = {\rho}(r)\exp(-|z|/\lambda)$, where $\lambda$ represents the vertical half-width of the source distribution, $z$ is the height above the GP and ${\rho}(r)$ is the source distribution at $z=0$ depending on Galactic radius $r$. 
The angular-integrated quasi-diffuse flux can then be expressed as:
\begin{equation}\label{eq:diffuse_formula}
\Phi_{\rm QD}(E_\nu) \equiv \int{d}\Omega\phi_{\rm QD}(E_\nu,\Omega) = Q_{\nu}(E_\nu)\Sigma_\odot\xi_{\rm gal}\,,
\end{equation}
where we introduced the local source surface density $\Sigma_\odot \equiv 2\lambda\rho(r_\odot)$ and the dimensionless $\mathcal{O}(1)$ parameter:
\begin{equation}
\xi_{\rm gal} \equiv \frac{1}{4\pi{\Sigma_\odot}}\int{d}\Omega\int_0^\infty{d} D\rho({\bf r}_\odot + D{\bf n}(\Omega))\,.
\end{equation}
We consider radial source distributions of the form~\cite{Green:2015isa}
\begin{equation}\label{eq:galactic_density}
\rho(r) = \rho_\odot \left(\frac{r}{r_{\odot}}\right)^{\alpha} e^{-\beta (r/r_{\odot}-1)}\,,
\end{equation}
where $\rho_\odot$ represents the (azimuthally averaged) local density in the solar neighborhood. 
As our benchmark case, we choose the distribution of SNRs analyzed in Ref.~\cite{Green:2015isa} with best-fit values $\alpha = 1.09 $, $\beta = 3.87$ and a vertical scale height $\lambda = 83\, \rm{pc}$ resulting in a Galactic form factor of $\xi_{\rm gal} \simeq 3$. We have investigated the dependence of our results by varying $\alpha$ and $\beta$ in Eq.~(\ref{eq:galactic_density}) within $[0,2]$ and $[1,4]$, respectively, for $\lambda=0.1$~kpc. We find that for this set of parameters the Galactic form factor varies only within $2.6\lesssim\xi_{\rm gal}\lesssim4.1$, indicating the robustness of our results.

In the following, we will focus on the contribution of unresolved sources with hard emission spectra ($\gamma=2$) that can contribute significantly to the soft Galactic diffuse spectrum ($\gamma\simeq2.5-2.7$) at the highest energies. As a pivot energy we choose $E_\nu=100$~TeV where IceCube's best-fit Galactic diffuse flux is at the level of $E_\nu^2\Phi \simeq 2\cdot10^{-8}~{\rm GeV}{\rm cm}^{-2}{\rm s}^{-1}$ independent of the Galactic emission models considered in Ref.~\cite{IceCube2023galactic} (see also App.~\ref{app:diffuse}). Figure~\ref{fig:density_luminosity} shows the source populations in terms of the monochromatic neutrino luminosity at 100~TeV defined as: 
\begin{equation}\label{eq:L100}
L_{100\,{\rm TeV}} \equiv [E_\nu^2Q_\nu(E_\nu)]_{E_\nu=100\,{\rm TeV}}\,,
\end{equation}
and the local source surface density $\Sigma_\odot$ (left axis) related to the \textit{expected}\footnote{Note that the actual number of sources could be significantly impacted by Poisson fluctuations in the case of low $\mathcal{N}$. This is not accounted for in this study.} total number of sources $\mathcal{N}$ (right axis). The green lines show the combinations of $L_{100\,{\rm TeV}}$ and $\Sigma_\odot$ that contribute to the observed angular-integrated Galactic neutrino emission at 100~TeV at levels of 1\%, 10\% and 100\%.

\section{Limits on Galactic Populations}\label{sec3}

The non-observation of individual Galactic neutrino sources by IceCube implies a limit on the Galactic source surface density $\Sigma_\odot$ and luminosity $L_{100~{\rm TeV}}$. We make use of IceCube's discovery potential (DP) $\Phi_{\rm DP}$ (units of ${\rm GeV}^{-1}\,{\rm cm}^{-2}\,{\rm s}^{-1}$) for point-like neutrino sources using track~\cite{IceCube:2019cia} and cascade events~\cite{IceCube2023galactic} that strongly depend on neutrino energy $E_\nu$ and source declination $\delta$. For a given source luminosity $L_{100\,{\rm TeV}}$ these discovery potentials define a declination-dependent discovery horizon of the form:
\begin{equation}
D_{\rm max}(\delta) \equiv \sqrt{\frac{L_{100\,{\rm TeV}}}{4\pi[E_\nu^2\Phi_{\rm DP}(E_\nu,\delta)]_{E_\nu=100\,{\rm TeV}}}}\,.
\label{eq:Dmax}
\end{equation}
Figure~\ref{fig:DP_MW} shows this horizon for Galactic sources for two IceCube analyses (``IC Tracks''~\cite{IceCube:2019cia} and ``IC Cascades''~\cite{IceCube2023galactic}) and a monochromatic neutrino luminosity $L_{100\,{\rm TeV}} = 10^{34}~{\rm erg}\,{\rm s}^{-1}$ as thick solid contours. We also indicate nearby potential neutrino sources from three source classes: SNRs and PWNe from the catalog search of Ref.~\cite{IceCube2023galactic} and a list of nearby YMSCs~\cite{Pfalzner:2009vn,Morales,Kharchenko:2005fe,Alvarez:2013ig} (see App.~\ref{app:sources} for details). The point-source DP of track events shows a particularly strong dependence on Galactic longitude related to the strong background of muons produced by CR interactions above the detector. Due to IceCube's location at the South Pole, where the zenith angle $\theta$ is degenerate with declination $\delta$ as $\theta = \delta +\pi/2$, this background affects the DP for sources in the Southern Sky, including sources in the direction of the GC. In contrast, the point-source DP of cascade events used in the study~\cite{IceCube2023galactic} has a more uniform coverage in terms of declination.

Note that the discovery horizons shown in Fig.~\ref{fig:DP_MW} assume point-like sources and have to be corrected for the enlarged angular extension of nearby sources. Assuming an (effective) source radius $R_{\rm src}$ and distance $D>R_{\rm src}$, the source angular radius becomes $\sigma_{\rm src} = \sin^{-1}(R_{\rm src}/D)$. We assume then that the DP of extended sources can be approximated as:
\begin{equation}\label{eq:scaledDP}
\Phi_{\rm DP}(E_\nu,\delta,\sigma_{\rm src}) \simeq \sqrt{\frac{\sigma^2_{\rm PSF}+\sigma_{\rm src}^2}{\sigma^2_{\rm PSF}}} \, \Phi_{\rm DP}(E_\nu,\delta)\,,
\end{equation}
where $\sigma_{\rm PSF}$ is the size of the point-spread function (PSF); see {\it e.g.}~Ref.~\cite{Steppa:2020qwe}. While this parameter in general depends on source declination and neutrino energy, we will use $\sigma_{\rm PSF}\simeq0.2^\circ$ ($\sigma_{\rm PSF}\simeq7^\circ$) for track (cascade) events at 100~TeV~\cite{IceCube:2019cia, IceCube2023galactic, KM3NeT:2018wnd, Gen2_TDR}.
Note that these angular resolutions represent optimistic values of the data samples that lead to conservative DP estimates from Eq.~(\ref{eq:scaledDP}).

\begin{figure}[t!]
\includegraphics[width=\linewidth]{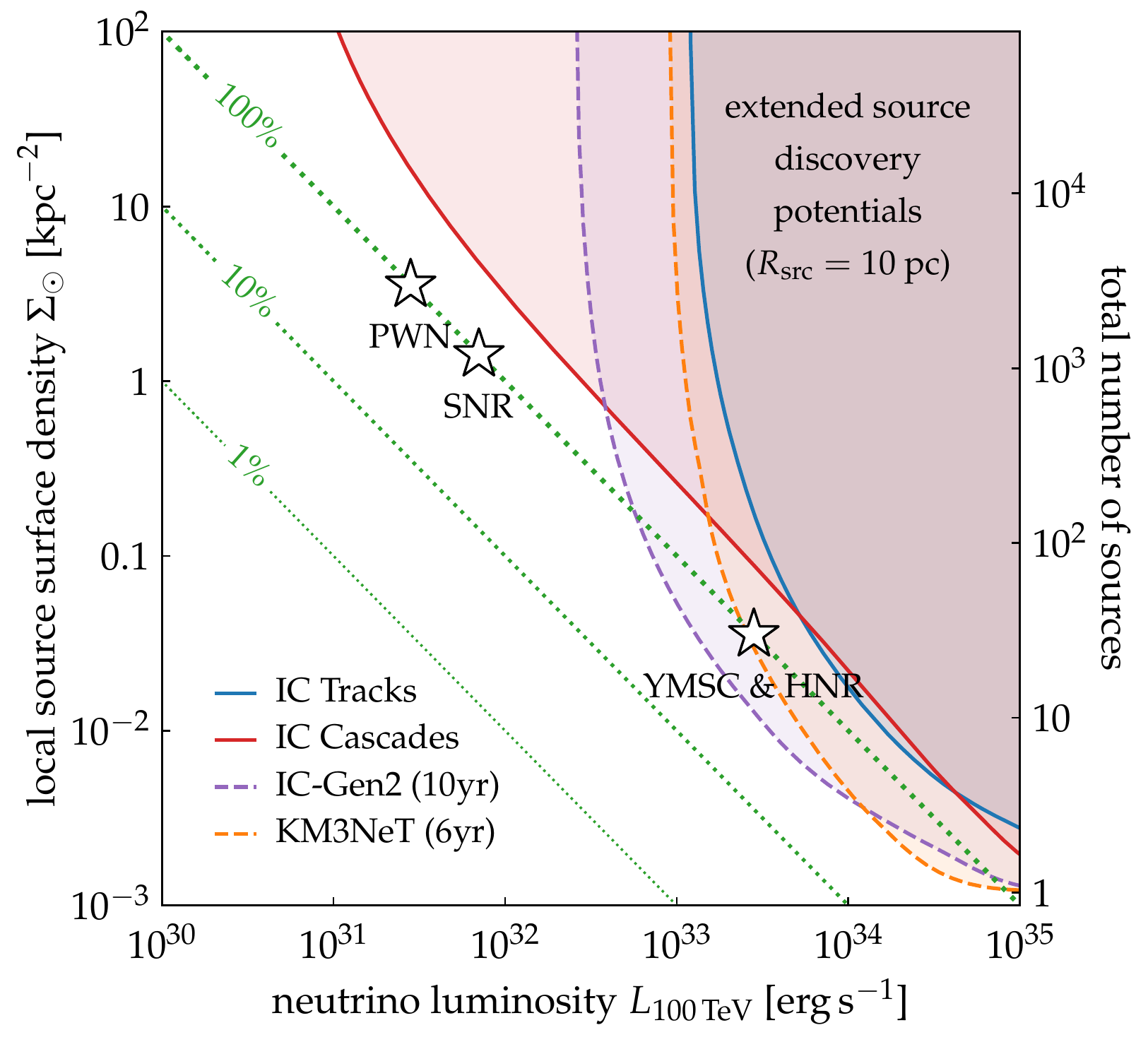}
\caption[]{
Comparison of the Galactic diffuse neutrino emission to the effective local surface density and luminosity of Galactic neutrino source populations. The green dashed lines show the contributions in terms of the observed angular-integrated neutrino flux at 100~TeV. The solid contours indicate populations where bright sources with an extension of $R_{\rm src}=10$~pc should have been discovered in IceCube’s point-source studies (``IC Tracks''~\cite{IceCube:2019cia} and ``IC Cascades''~\cite{IceCube2023galactic}). The dashed contours show the expected reach of KM3NeT~\cite{KM3NeT:2018wnd,KM3NeT:2021szv} and the proposed IceCube-Gen2 facility~\cite{IceCube-Gen2:2020qha,Gen2_TDR}. We also indicate the required luminosity of pulsar wind nebulae (PWNe), supernova remnants (SNRs), hypernovae remnants (HNRs) and young massive star clusters (YMSCs) to saturate the diffuse emission at 100~TeV.}\label{fig:density_luminosity}
\end{figure}

We can now evaluate the expected number $N_{\rm obs}$ of observed sources as: 
\begin{equation}\label{eq:number_sources_IceCube}
N_{\rm obs} = \int{d}\Omega \int_{R_{\rm src}}^{D_{\rm max}(\delta)} {d}D D^2 \rho({\bf r}_\odot + D{\bf n}(\Omega))\,,
\end{equation}
where $D_{\rm max}(\delta)$ accounts for the scaled DP of Eq.~(\ref{eq:scaledDP}). So far, no Galactic neutrino point sources have been identified, which implies an upper limit $N_{\rm obs} \lesssim 1$. Figure~\ref{fig:density_luminosity} shows the corresponding exclusion limits of neutrino sources using IC tracks (solid blue contour) and IC cascades (solid red contour). We assume here that the sources have an extension of $R_{\rm src} = 10$~pc, motivated by the typical size of a SNR at the end of the Sedov-Taylor phase~\cite{Reynolds2008}.
Interestingly, IceCube's current source DPs are not sufficient to exclude a $100\%$ contribution to the Galactic diffuse flux over a wide range of source surface densities and luminosities. Figure~\ref{fig:density_luminosity} also indicates the required luminosities for different source types examined in detail in Section~\ref{sec4}.

Figure~\ref{fig:DP_MW} also shows the expected discovery horizon for KM3NeT ARCA~\cite{KM3Net:2016zxf} as well as the planned IceCube-Gen2~\cite{IceCube-Gen2:2020qha} (using the 10 year DP with surface array) for the same benchmark luminosity. Using track events, optical Cherenkov telescopes in the Northern Hemisphere are expected to have an increased discovery horizon for sources towards the GC. Notably, a recent analysis by ANTARES~\cite{ANTARES:2022izu} finds a hint for TeV neutrino emission from the Galactic Ridge, although with weak significance and consistent with earlier upper limits~\cite{ANTARES:2018nyb}. The expected exclusion contours of KM3NeT and IceCube-Gen2 are shown in Fig.~\ref{fig:density_luminosity} as dashed contours. These detectors will be able to probe the contribution of rare but powerful Galactic sources if they dominate ($>50\%$) the diffuse emission at 100~TeV as long as the source extension is limited to about $10$~pc.

Note that, to be conservative, the KM3NeT DP from Ref.~\cite{KM3NeT:2018wnd} shown in Fig.~\ref{fig:DP_MW} excludes the region $\delta \gtrsim 50^{\circ}$ which is only visible above the horizon~\cite{KM3Net:2016zxf,KM3NeT:2018wnd}. However, similar to IceCube, future event selections of KM3NeT are also expected to probe neutrino sources via high-energy track events at high declination angles. Likewise, KM3NeT is also expected to have a good sensitivity and angular resolution to cascade events~\cite{KM3Net:2016zxf}; see also Ref.~\cite{Sudoh:2023qrz}. Similarly, IceCube-Gen2 is also expected to improve the detection prospects of Galactic neutrino sources with the inclusions of cascade events as well as by a surface veto for atmospheric background events~\cite{IceCube-Gen2:2020qha,Gen2_TDR}.

\begin{figure*}[t!]
\includegraphics[width=0.49\linewidth]{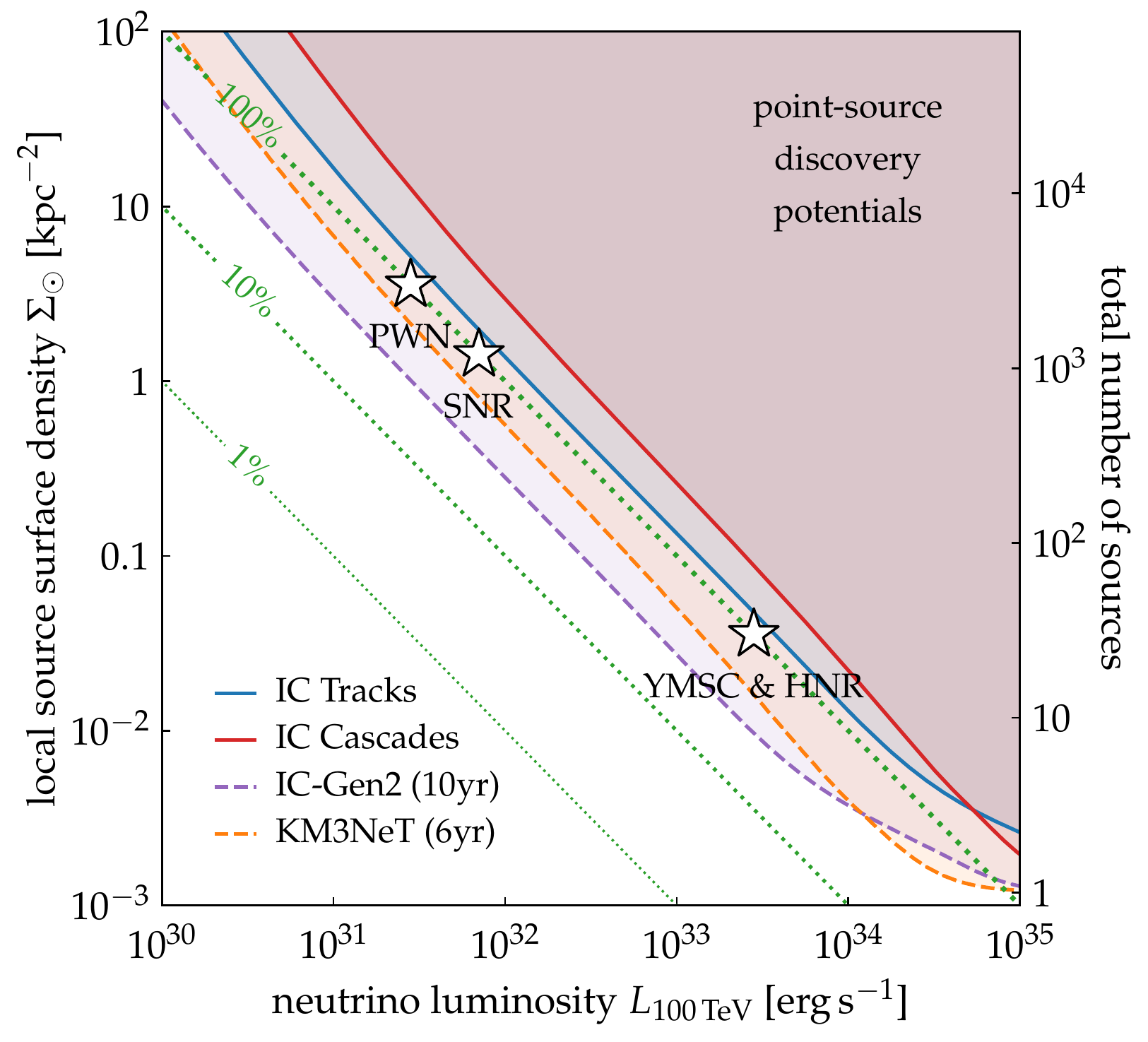}\hfill\includegraphics[width=0.49\linewidth]{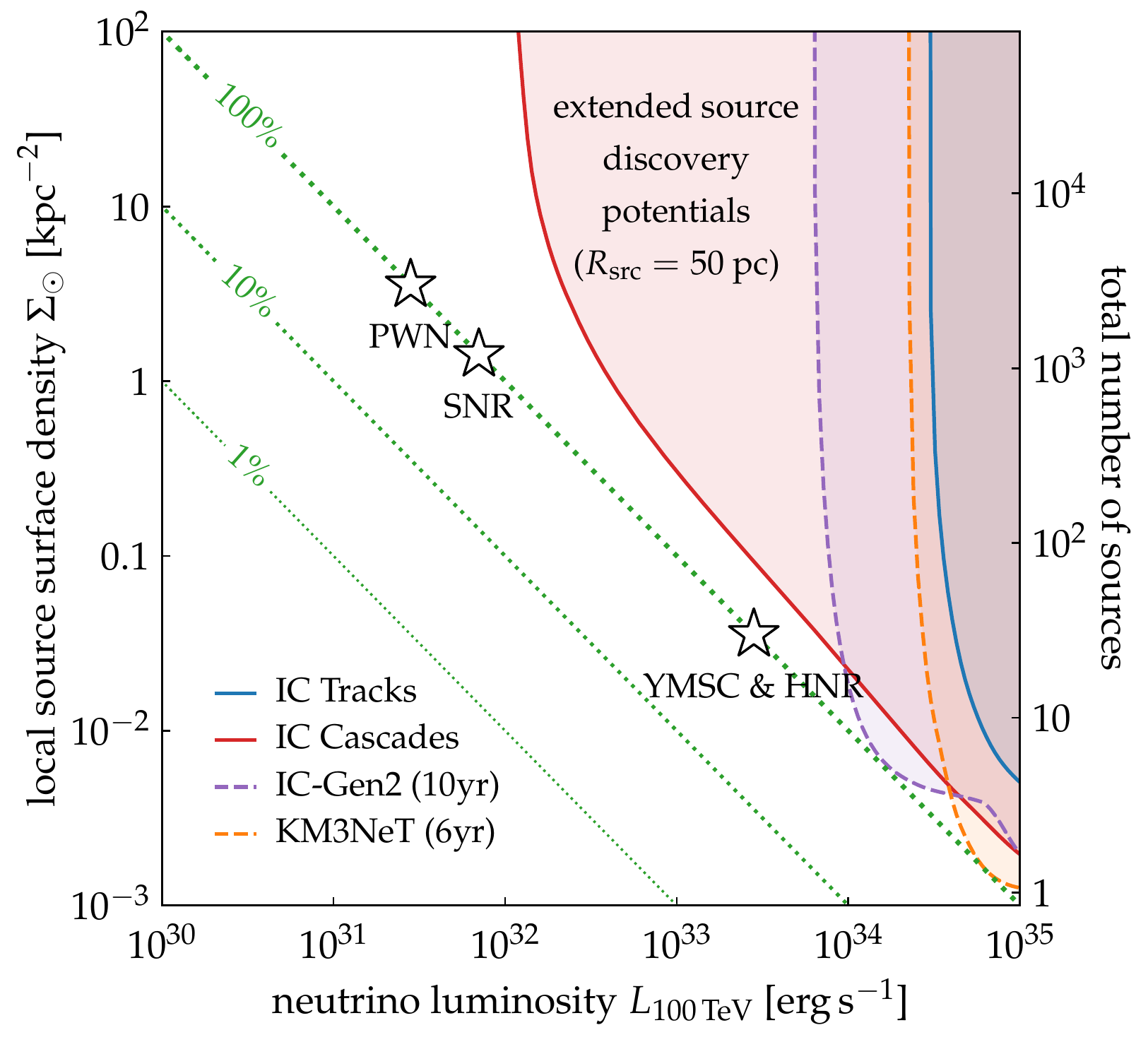}
\caption[]{Same as Fig.~\ref{fig:density_luminosity} but now showing the discovery potential for point-like sources (left panel) and for sources with a 50~pc radius (right panel).}
\label{fig:SNR_50pc}
\end{figure*}

The discovery horizon of Galactic sources depends strongly on the source extension.
As an illustration, Fig.~\ref{fig:SNR_50pc} shows the exclusion limits of Galactic populations (as compared to Fig.~\ref{fig:density_luminosity}) for point-like sources (left panel) and sources with a radius of $R_{\rm src}=50$~pc (right panel), typical for YMSCs~\cite{HESS:2022pwf} and also an average value for the radius of a Pulsar TeV Halo, which can extend up to $\sim 100\, \rm{pc}$~\cite{Linden:2017vvb}. Indeed, identifying PeVatrons of large extension will be challenging for the upcoming detectors, even though dedicated multi-messengers analyses might improve the discovery prospects. Note that the source extension is less relevant for the cascade-based analyses of Ref.~\cite{IceCube2023galactic} due to the large intrinsic angular uncertainty of event reconstructions in IceCube. We also emphasize that in a more realistic scenario sources will have different sizes, and this could impact the limits as well. For instance, if local sources have a reduced radius with respect to sources near the Galactic Center, this might well lead to limits very near to the point-like case shown in the left panel of Fig.~\ref{fig:SNR_50pc}.

IceCube also searched for the combined neutrino emission from three catalogs of SNRs, PWNe, and unidentified $\gamma$-ray sources in Ref.~\cite{IceCube2023galactic}, updating previous stacking searches in IceCube \cite{IceCube:2019cia,Liu:2019iga}. Each catalog was comprised of 12 local $\gamma$-ray sources with most promising expectations for neutrino emission under the hypothesis of correlated $\gamma$-ray and neutrino production from CR interactions. Assuming an equal weight for each source, the IceCube analysis finds an excess of more than $3\sigma$ from each of these catalogs; however, as already pointed out in Ref.~\cite{IceCube2023galactic}, it is difficult to interpret these results as independent evidence of neutrino sources due to the spatial overlap with the Galactic diffuse emission templates and the limited angular resolution of the cascade data. 

We will therefore consider in the following the per-flavor upper limits of IceCube's catalog stacking searches to derive bounds on the monochromatic neutrino luminosity of SNRs and PWNe. At the $90\%$ confidence level, the combined flux is limited to $E_\nu^{2} \Phi^{90\% {\rm UL}}_{\rm stack} \simeq 9.0\cdot 10^{-9}\, \rm{GeV}\, \rm{cm}^{-2}\, \rm{s}^{-1}$ for the SNR catalog, and $E_\nu^{2} \Phi^{90\% {\rm UL}}_{\rm stack} \simeq 9.5\cdot 10^{-9}\, \rm{GeV}\, \rm{cm}^{-2}\, \rm{s}^{-1}$ for PWNe. Assuming a constant prior for the source distance $D^{\rm min}_i<D<D^{\rm max}_i$, we can obtain an upper limit on the neutrino luminosity of $L_{100\,{\rm TeV}} < 3.7\cdot 10^{32} \, \rm{erg}\,\rm{s}^{-1}$ for SNRs and $L_{100\,{\rm TeV}}< 1.2\cdot 10^{32} \, \rm{erg} \,\rm{s}^{-1}$ for PWNe. We refer to App.~\ref{app:sources} for details and source catalogs. We will discuss the relevance of these results in the next section.

\section{Candidate Neutrino Sources}\label{sec4}

CRs in our Galaxy produce neutrinos via collisions with gas. The cross section for inelastic proton-proton interactions is $\sigma_{\rm pp} \simeq (3-6)\cdot10^{-26}\,{\rm cm}^2$ for TeV--PeV CR protons with mean inelasticity of $\kappa_p\simeq0.5$~\cite{Kelner:2006tc}. The interaction rate of PeV CR protons is then $t_{\rm pp} = (\sigma_{\rm pp}cn_{\rm gas})^{-1} \simeq 17.6\,{\rm Myr}(n_{\rm gas}/n_{\rm ISM})^{-1}$ where $n_{\rm gas}$ is the target gas density and $n_{\rm ISM} \simeq 1{\rm cm}^{-3}$ the typical gas density in the interstellar medium (ISM). The CR energy loss is dominated by pion production with nearly equal contributions of $\pi^+$, $\pi^0$ and $\pi^-$ due to isospin symmetry. Charged pions $\pi^\pm$ produced in these collisions subsequently decay to neutrinos via $\pi^+\to \mu^++\nu_\mu$ followed by $\mu^+ \to e^++\nu_e+\bar\nu_\mu$ and the charge-conjugate processes. The average neutrino energy from these decays is $E_\nu \simeq 0.05E$. Accounting for neutrino flavor oscillations, we can then estimate the per-flavor neutrino spectral production rate (units of ${\rm GeV}^{-1}\,{\rm s}^{-1}$) as $E_\nu^2Q_{\nu}\simeq(\kappa_{\rm pp}/6)t^{-1}_{\rm pp}E^2N_{\rm CR}$ where $N_{\rm CR}(E)$ is the CR spectrum (units of ${\rm GeV}^{-1}$).

\subsection*{Diffuse Emission in the Interstellar Medium}

Galactic diffuse neutrino emission is produced by collisions of CRs with the ISM as they propagate through the Milky Way. The local CR density is $E^2n_{\rm CR}(E) \simeq 7.5\cdot10^{-10} (E/{\rm GeV})^{-0.7}~{\rm GeV}\,{\rm cm}^{-3}$ for CRs in the GeV--PeV energy region~\cite{ParticleDataGroup:2022pth}. The neutrino flux is therefore expected to inherit the soft spectrum of local CRs with $Q_{\nu}\propto E_\nu^{-2.7}$. While CRs are expected to diffuse several kpc out of the GP, neutrino production will be dominated by their interactions with the relatively dense ISM within $|z|\leq \lambda\simeq0.1$~kpc of the disk~\cite{Johannesson:2018bit}. Using the notation of section~\ref{sec2}, we can derive the neutrino luminosity per surface area from the CR surface density $\Sigma_\odot N_{\rm CR}\simeq 2\lambda n_{\rm CR}$. For CR protons with $E \simeq 20E_\nu = 2$~PeV, we arrive at $\Sigma_\odot L_{100\,{\rm TeV}}\simeq  4\cdot10^{31} \, {\rm erg}\,{\rm s}^{-1}\,{\rm kpc}^{-2}$ which yields an angular-integrated diffuse flux of:
\begin{equation}\label{eq:ISMdiffuse}
E_\nu^2\Phi_{\rm MW} \simeq 8\cdot10^{-9}\left(\frac{\xi_{\rm gal}}{3}\right)\left(\frac{E_\nu}{100~ {\rm TeV}}\right)^{-0.7}\frac{{\rm GeV}}{{\rm cm}^{2}{\rm s}}\,.
\end{equation}
Note that this is only an order-of-magnitude estimation assuming a pure proton CR flux and not accounting for spectral breaks. More detailed calculations account for a heavier CR composition in the CR {\it knee} region as well as non-uniform CR distributions~\cite{Joshi:2013aua,Kachelriess:2014oma,Ahlers:2015moa,Gaggero:2015xza,Schwefer:2022zly,DelaTorreLuque:2022ats}.
However, our estimate (\ref{eq:ISMdiffuse}) is indeed in the vicinity of the flux level inferred by IceCube (see Fig.~\ref{fig:diffuse_fluxes}) and indicates the strong contribution of Galactic diffuse emission to the signal. 

\subsection*{Supernova Remnants}

QD neutrino emission associated with CR interactions near or within their sources is expected to follow the hard emission spectrum $Q_{\nu}\propto E_\nu^{-2}$ of freshly accelerated CRs. SNRs have been argued to supply the bulk of Galactic CRs below the CR {\it knee}~\cite{Baade:1934zex,1961PThPS..20....1G}. By the end of the Sedov-Taylor phase, a significant energy fraction $\epsilon_{\rm CR}$ of the initial supernova (SN) ejecta energy of ${\mathcal E}_{\rm ej}$ could be transferred to CRs via diffusive shock acceleration (DSA)~\cite{1977DoSSR.234.1306K,Blandford:1978ky,1977ICRC...11..132A,Bell:1978zc,Bell:1978fj} giving  $E^2N_{\rm CR} \simeq \epsilon_{\rm CR}\chi{\mathcal E}_{\rm ej}$ with $\chi \simeq 1/\ln({\rm PeV}/{\rm GeV}) \simeq 0.07$. For typical ejecta energies of $\mathcal{E}_{\rm ej} \simeq 10^{51}$~erg expanding into the ISM we then arrive at the estimate $L_{100\,{\rm TeV}} \simeq \epsilon_{\rm CR}10^{34}(\mathcal{E}_{\rm ej}/10^{51}{\rm erg}) \,{\rm erg}\,{\rm s}^{-1}$. Interestingly, the upper limit from the SNR catalog search limits the CR acceleration efficiency to a level of about 1\% for our choice of benchmark source parameters. CR acceleration remains effective until the onset of the radiative phase at $t_{\rm RP}\simeq4\cdot{10}^{4}\,{\rm yr}\,({\mathcal E}_{\rm ej}/10^{51} \, {\rm erg})^{4/17}(n_{\rm gas}/n_{\rm ISM})^{-9/17}$~\cite{1998ApJ...500..342B}. We can then estimate the number of active SNRs from the local SN rate of $R_{\rm SN}\sim0.03~{\rm yr}^{-1}$ as $\mathcal{N}_{\rm SNR}\simeq R_{\rm SN}t_{\rm RP}\simeq1200$ \cite{Ahlers:2013xia}, which corresponds to a local surface source density of $\Sigma_\odot \simeq 1.6~{\rm kpc}^{-2}$. Based on this surface density we can estimate that $L_{100\,{\rm TeV}}\simeq 6\cdot10^{31}\,{\rm erg}\,{\rm s}^{-1}$ would be sufficient to saturate the diffuse flux, as can be seen in Fig.~\ref{fig:density_luminosity}. This is consistent with the above estimate of the neutrino luminosity for moderate CR acceleration efficiencies.

Note that, for the production of 100~TeV neutrinos, SNRs have to be capable of reaching CR energies at the level of PeV, which poses a challenge for standard DSA theory; see {\it e.g.}~the reviews~\cite{Blasi:2013rva,Amato:2017dbs,Gabici:2019jvz}. The conditions are more favorable for powerful, $\mathcal{E}_{\rm ej}\simeq10^{52}$~erg, but rare core-collapse SNe with relatively small ejecta mass~\cite{Sveshnikova:2003sa,Wang:2007ya}. The rate of these hypernovae is $\sim1\textrm{--}2$\% of the SN rate~\cite{Guetta:2006gq,Arcavi:2010wd,Cristofari:2020mdf} corresponding to about $\mathcal{N}_{\rm HNR}\simeq30$ active hypernova remnants (HNRs) and a local surface source density of $\Sigma_\odot \simeq 0.04~{\rm kpc}^{-2}$. The required luminosity to saturate the diffuse flux becomes now $L_{100\,{\rm TeV}}\simeq 2.5\cdot10^{33}\,{\rm erg}\,{\rm s}^{-1}$ as shown in Fig.~\ref{fig:density_luminosity} and is consistent with our earlier estimate for moderate CR acceleration efficiencies of a few percent.

\subsection*{Pulsar Wind Nebulae}

There is a general consensus in ascribing the observed multiwavelength radiation from PWNe to leptonic mechanisms~\cite{Bucciantini:2010pd}. However, protons could also be extracted from the surface of spinning neutron stars and accelerated up to very high energies~\cite{Kotera:2015pya}. They could then be confined for a long time in a region surrounding the nebula, often referred to as a TeV halo, characterized by reduced diffusion~\cite{Mukhopadhyay:2021dyh}. The total number of $\gamma$-ray emitting PWNe in the Galaxy can be estimated from the SN rate $R_{\rm SN}$ and the lifetime $t_{\rm PWN} \lesssim 10^5\,{\rm yr}$ as $\mathcal{N}_{\rm PWN} \simeq R_{\rm SN} t_{\rm PWN} \simeq 3000$ \cite{Fiori:2022nhu}, corresponding to a local surface density of $\Sigma_{\odot} \simeq 3.6\,{\rm kpc}^{-2}$. The corresponding neutrino luminosity required to saturate the diffuse flux becomes $L_{100\,{\rm TeV}}\simeq 2.8\cdot10^{31}\,{\rm erg}\,{\rm s}^{-1}$ as shown in Fig.~\ref{fig:density_luminosity}. Considering an initial spin-down power $L_0$ we estimate the time-integrated CR spectrum as $E^2N_{\rm CR} \simeq \epsilon_{\rm CR}\eta t_{\rm PWN}L_0\simeq \epsilon_{\rm CR}2\cdot10^{50}\,{\rm erg}$ where $\epsilon_{\rm CR}$ is the conversion efficiency of spin-down power into CRs. The neutrino luminosity can now be estimated as $L_{100 \rm TeV} \simeq (\kappa_{\rm pp}/6)t^{-1}_{\rm pp}(t_{\rm diff}/t_{\rm PWN})E^2N_{\rm CR} \simeq \epsilon_{\rm CR} ~3\cdot10^{32}{\rm erg}\,{\rm s}^{-1}$, where we estimate $t_{\rm diff}\simeq 10^3\,{\rm yr}$ as the timescale of diffusive PeV CR escape from the PWN. Therefore, considering high CR acceleration efficiencies, it is feasible that PWNe can contribute substantially to the Galactic diffuse flux.

\subsection*{Young Massive Star Clusters}

YMSCs are another class of promising candidate CR PeVatrons~\cite{Morlino:2021xpo}. They are often defined as compact stellar aggregates characterized by a cluster mass $\gtrsim 10^4 \, \rm M_{\odot}$ and a relatively young age $t_{\rm YMSC} \sim 1-10\,\rm{Myr}$~\cite{PortegiesZwart:2010cly}. YMSCs can sustain powerful winds of kinetic luminosity $L_{\rm kin} \sim 10^{38}-10^{39} \, \rm erg \, s^{-1}$ where CRs can be efficiently accelerated at the wind termination shock~\cite{Morlino:2021xpo}. In addition, SNe can dominate the CR injection power $5\textrm{--}10$~Myr after the cluster formation~\cite{Vieu:2022exk}. The local number density of YMSCs can be estimated from the local star formation rate density $\sim 5000 \, \rm M_{\odot} \, Myr^{-1} \, kpc^{-2}$. Assuming that about $10\%$ of stars end up in bound star clusters and that YMSCs contribute about $10\%$ to the overall star cluster luminosity function (see Ref.~\cite{PortegiesZwart:2010cly} for details) it is possible to estimate the YMSC surface density as $\Sigma_{\odot} \sim 0.02\textrm{--}0.2\,{\rm kpc}^{-2}$ for which one can expect diffusive shock acceleration at the wind termination shock to be the dominant acceleration mechanism. For $\Sigma_{\odot} \simeq 0.04\,{\rm kpc}^{-2}$, that we also used in our previous estimate for HNRs, we see that the luminosity required to saturate the diffuse flux becomes $L_{100\,{\rm TeV}}\simeq 2.5\cdot10^{33}\,{\rm erg}\,{\rm s}^{-1}$ as seen in Fig.~\ref{fig:density_luminosity}.  
Combining this result with the YMSC catalog of App.~\ref{app:sources}, the total expected neutrino flux of the YMSCs in the catalog is of the order $E_\nu^{2} \Phi^{90\% {\rm UL}}_{\rm stack} \simeq 1.6\cdot 10^{-8}\, \rm{GeV}\, \rm{cm}^{-2}\, \rm{s}^{-1}$.
Assuming CR acceleration in the wind termination shock~\cite{Morlino:2021xpo}, we can estimate the time-integrated CR spectrum as $E^2 N_{\rm CR} = \epsilon_{\rm CR} \chi t_{\rm YMSC}L_{\rm kin} \simeq \epsilon_{\rm CR}2\cdot10^{51}$~erg for typical YMSC conditions. The neutrino luminosity can now be estimated as $L_{100 \, \rm TeV} \simeq (\kappa_{\rm pp}/6)t^{-1}_{\rm pp}E^2N_{\rm CR} \simeq \epsilon_{\rm CR}~ 3\cdot10^{34}~{\rm erg}\,{\rm s}^{-1}$, where we assumed $n_{\rm gas} \simeq 0.1n_{\rm ISM}$ as the average gas density in the shocked wind region. Again, for high CR acceleration efficiencies, it is feasible that the neutrino emission from YMSC saturates the diffuse flux level.

\section{Conclusions}\label{sec5}

In this article, we have investigated the contribution of unresolved Galactic neutrino sources to the Galactic diffuse emission recently reported by IceCube. We have shown that the combined contribution of Galactic sources as a QD flux can be dominant at 100~TeV, which is consistent with the non-observation of individual Galactic neutrino sources so far. We argue that, due to IceCube's limited angular resolution of cascade events, the spatial morphology of QD emission is practically indistinguishable from diffuse emission from CR interactions in the ISM.

We have discussed the expected quasi-diffuse flux of unresolved source in a model-independent way by considering the average surface source density and luminosity of Galactic source populations. We have shown that our results are robust against variations of Galactic source distributions along the Galactic plane. Future observations of Galactic neutrino sources will allow us to improve our methods, {\it e.g.}~by modeling sources via their luminosity function and their variability of source spectra.

To estimate the present discovery horizon of individual Galactic neutrino sources, we have utilized IceCube's point-source DPs for $E^{-2}$ source spectra. We accounted for source extensions by rescaling the point-source DPs with the source solid angle, which is a good approximation for background-dominated data. We point out that realistic emission spectra of Galactic sources are expected to show breaks or cutoffs above 100~TeV inherited from the maximal CR energy achievable in Galactic sources. Dedicated IceCube analyses of extended Galactic neutrino sources will be able to provide us with a more realistic discovery horizon. However, we argue that even under the most optimistic conditions of unbroken power-law emission and negligible source extension, IceCube's discovery horizon for Galactic sources is presently insufficient to exclude a dominant QD emission of Galactic sources at 100~TeV.

We have also estimated the prospects of the upcoming KM3NeT ARCA detector and the planned IceCube-Gen2 facility to probe Galactic QD emission. For these estimates we assumed a fiducial source radius of 10~pc and rather conservative assumptions on the capability of these future detectors. The most promising targets for future studies are rare and luminous CR PeVatrons, with HNRs and YMSCs being excellent candidates. 

Finally, another powerful probe of Galactic neutrino sources is the associated emission of hadronic $\gamma$-rays~\cite{Cataldo:2020qla,Steppa:2020qwe,Vecchiotti:2021yzk, Vecchiotti:2023ill}. 
The monochromatic $\gamma$-ray luminosity is about twice the neutrino luminosity $L_{100~{\rm TeV}}$ that we used as one of our model parameters~\cite{Ahlers:2013xia}. This estimate does not account for absorption processes and additional $\gamma$-ray production from leptonic mechanisms, though. We will leave a discussion of this constraint to a future publication. We do however note that the measured diffuse Galactic neutrino flux by IceCube is consistent with the flux level inferred by $\gamma$-ray observatories in the TeV--PeV range such as Tibet and LHAASO \cite{IceCube2023galactic}. 
Therefore, even if the Galactic neutrino flux is dominated by unresolved sources at $E_{\nu} \sim 100\, \rm TeV$, their neutrino flux does not exceed the corresponding $\gamma$-ray data at those energies.

\begin{acknowledgments}
The authors would like to thank the IceCube Collaboration for valuable comments. A.A.~is supported by the research Grant No. 2017W4HA7S ``NAT-NET: Neutrino and Astroparticle Theory Network'' under the program PRIN 2017 funded by the Italian Ministero dell'Universit\`a e della Ricerca (MUR) and by the research project TAsP (Theoretical Astroparticle Physics) funded by the Istituto Nazionale di Fisica Nucleare (INFN). M.A., K.M.G., and E.P.~acknowledge support by Villum Fonden (No.~18994). E.P.~was also supported by the European Union’s Horizon 2020 research and innovation program under the Marie Skłodowska-Curie Grant Agreement No.~847523 ``INTERACTIONS''.
\end{acknowledgments}

\bibliographystyle{utphys_mod}
\bibliography{references}

\appendix

\begin{figure*}[p!]
\centering
\includegraphics[height=0.46\linewidth]{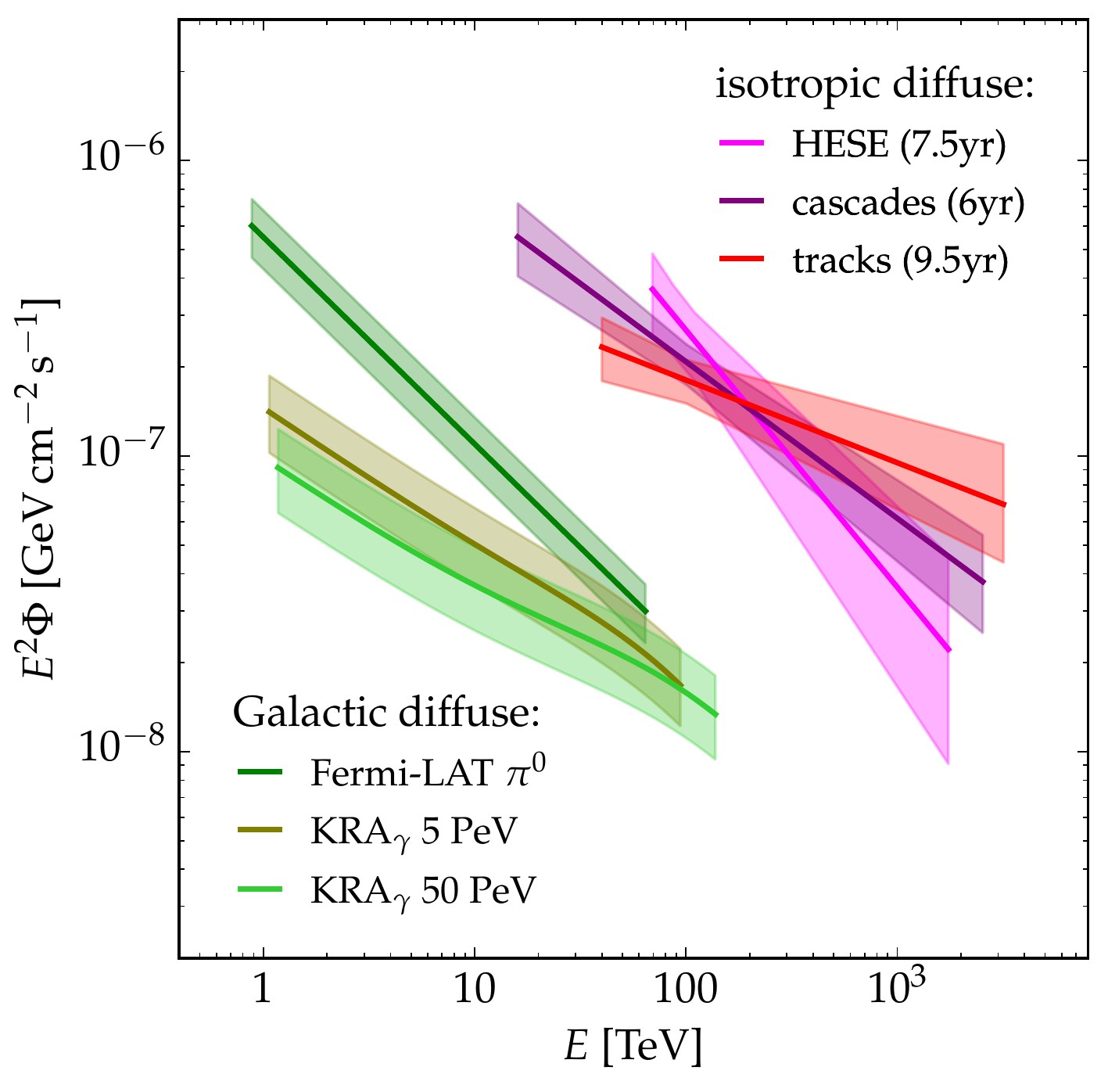}\hfill\includegraphics[height=0.46\linewidth]{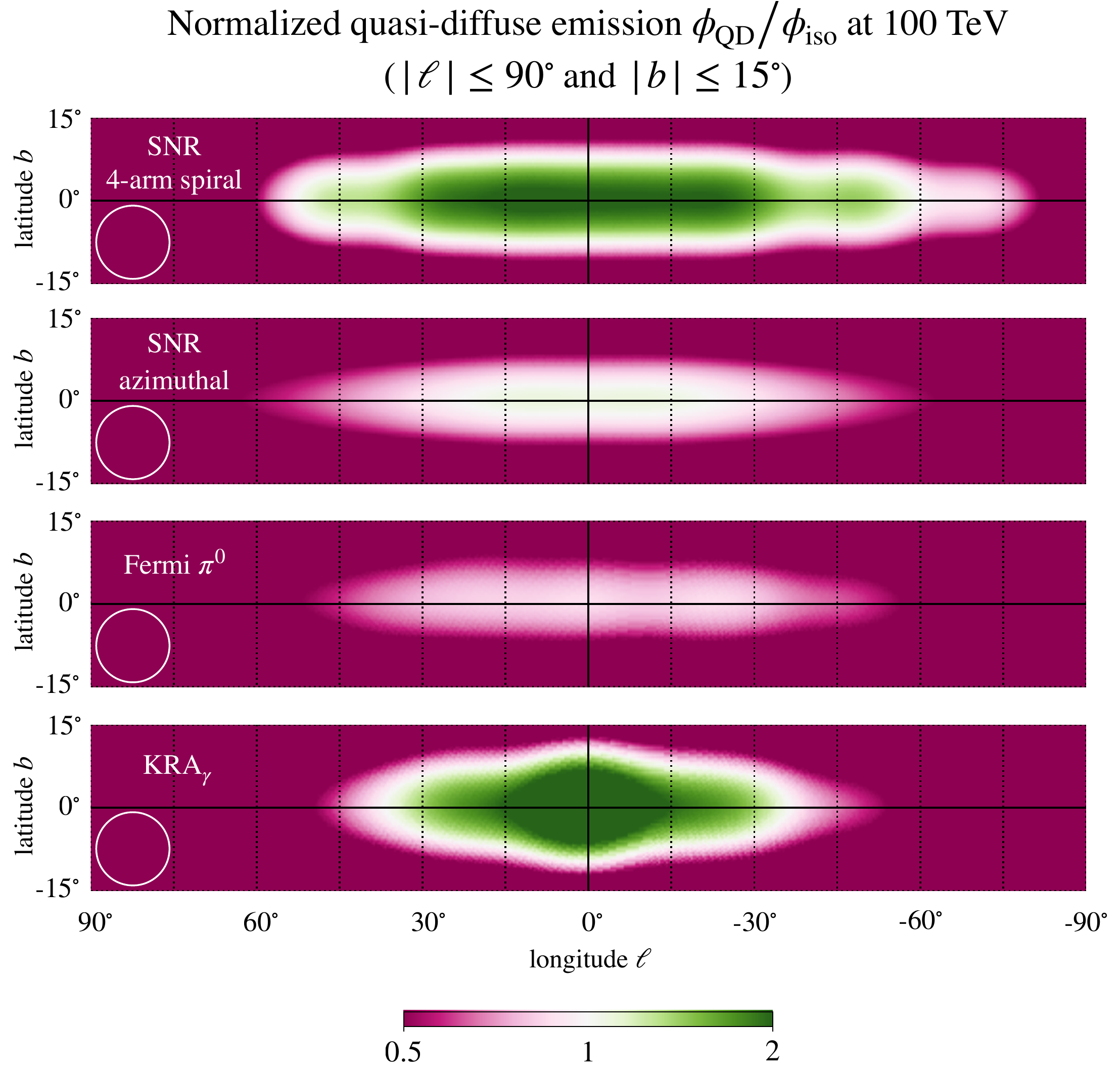}
\caption[]{{\bf Left panel:} Summary of IceCube's observations of isotropic and Galactic diffuse neutrino emission. The plot shows the angular-integrated diffuse flux $\Phi$ of isotropic emission (red bands: HESE~\cite{IceCube:2020wum}, cascades~\cite{IceCube:2020acn} and tracks~\cite{IceCube:2021uhz}) and Galactic emission (green bands: Fermi-LAT $\pi^0$~\cite{Fermi-LAT:2014ryh} and KRA${}_\gamma$~\cite{Gaggero:2015xza,GaggeroData}). The spectra are indicated by the best-fit spectrum (solid line) and the $1\sigma$ uncertainty range (shaded range). {\bf Right panel:} Comparison of diffuse and quasi-diffuse emission templates from the inner Galaxy. The template is smoothed over a Gaussian kernel with FWHM$=14^\circ$ (white circle) corresponding to typical angular resolution of $7^\circ$ of IceCube's cascade sample.}\label{fig:diffuse_fluxes}
\end{figure*}
\begin{figure*}[p!]
\centering
\includegraphics[height=0.46\linewidth,clip=true,bb = 0 60 630 655]{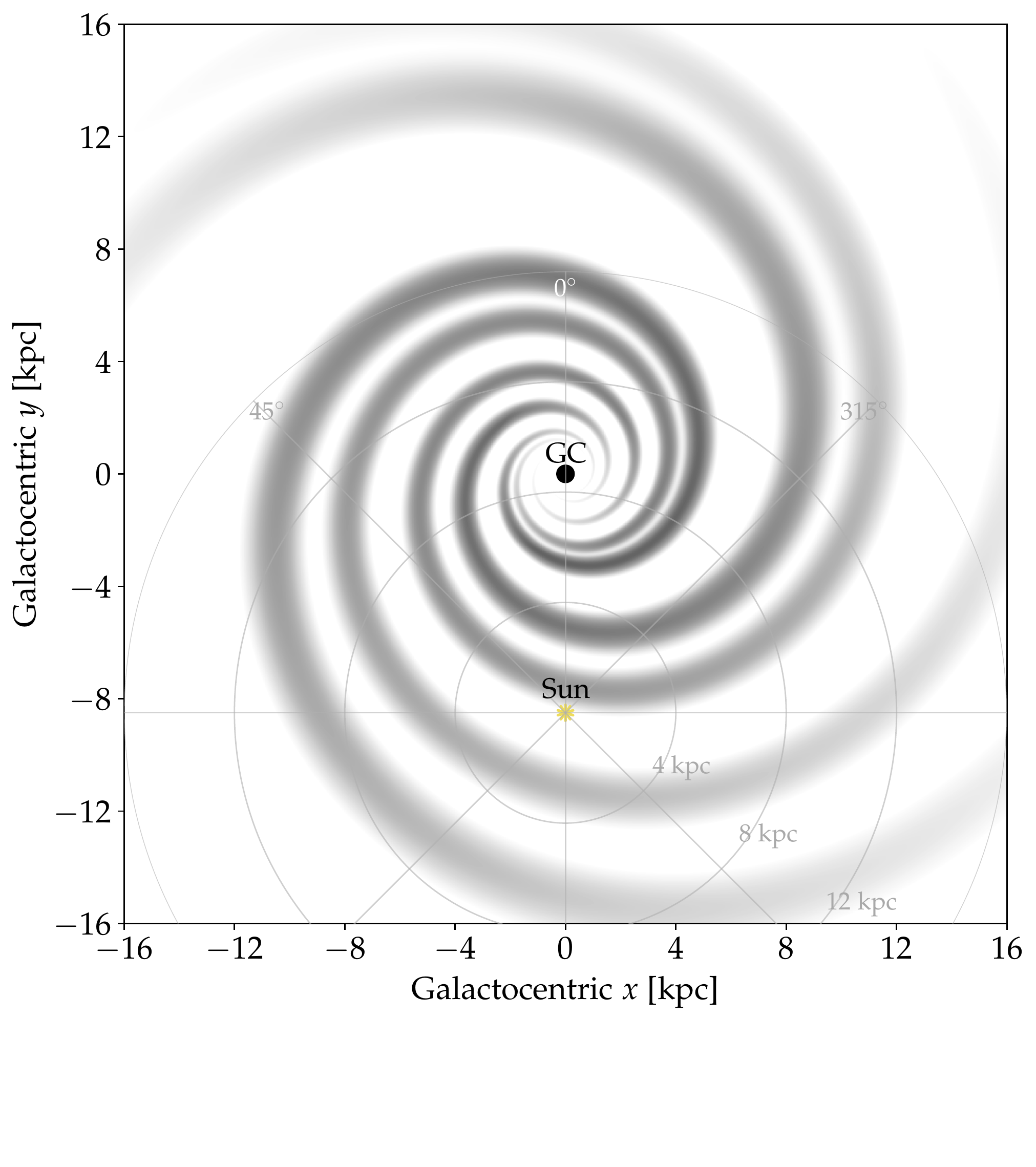}\hfill\includegraphics[height=0.46\linewidth]{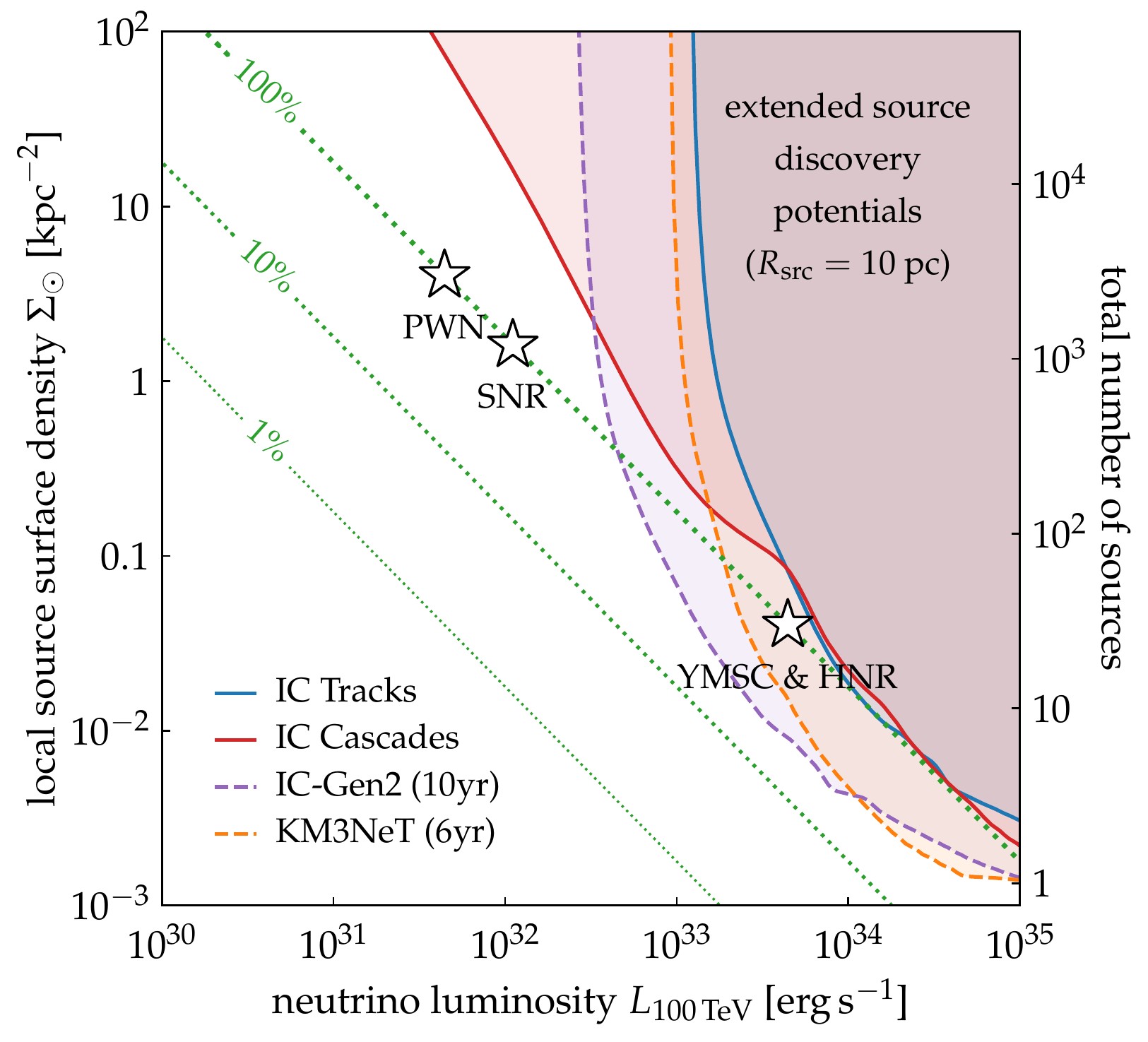}
\caption[]{{\bf Left panel:} The 4-arm spiral parametrization of Eq.~(\ref{eq:spiral}) superimposed over the radial SNR distribution from Ref.~\cite{Green:2015isa}. Note that for this visualization we rotated the Galactic coordinate system so that the Solar System appears at $x_\odot=0$ and $y_\odot=-8.5$~kpc. {\bf Right panel:} The corresponding quasi-diffuse flux and exclusion limits from the non-discovery of Galactic sources using the 4-arm model (as compared to Fig.~\ref{fig:density_luminosity}).}\label{fig:4arm}
\end{figure*}

\section{Diffuse Neutrino Fluxes}\label{app:diffuse}

In this section, we summarize IceCube's observations of diffuse neutrino fluxes and discuss the difference in the spatial templates of the Galactic diffuse and quasi-diffuse fluxes. The left panel of Fig.~\ref{fig:diffuse_fluxes} shows the spectral energy distributions of isotropic diffuse emissions (``HESE''~\cite{IceCube:2020wum}, ``cascades''~\cite{IceCube:2020acn} and ``tracks''~\cite{IceCube:2021uhz}) as well as the recent results of Galactic diffuse emission (Fermi-LAT $\pi^0$~\cite{Fermi-LAT:2014ryh} and KRA${}_\gamma$~\cite{Gaggero:2015xza} templates). The fluxes are integrated over solid angle for this comparison. The angular-integrated Galactic flux at 100~TeV is at the level of $\sim 10\%$ of the isotropic flux level. However, the Galactic diffuse flux is dominating over the isotropic emission along the Galactic Plane as indicated in the right panel of Fig.~\ref{fig:diffuse_fluxes}. Due to the limited angular resolution of the IceCube analysis of typically $7^\circ$, the predicted emission templates are smeared out using a Gaussian kernel with full-width half maximum of $14^{\circ}$ using {\tt HealPix} tools ~\cite{Gorski:2004by,Zonca:2019vzt}. The right panel shows, from top to bottom, the spatial distribution of the Galactic emission (normalised to the isotropic flux level at $100\, \rm{TeV}$) for the 4-spiral arm distribution (Eq.~\ref{eq:spiral}), the SNR-azimuthal distribution (Eq.~\ref{eq:galactic_density}), the Fermi-LAT $\pi^0$ template \cite{Fermi-LAT:2014ryh}, and the $\rm{KRA}_{\gamma}$ template~\cite{Gaggero:2015xza}. On can notice that all these templates predict an enhanced neutrino emission towards the inner Galaxy, with some variation in morphology and flux level. As already pointed out in Ref.~\cite{IceCube2023galactic}, the similarity of these templates makes the disentanglement of Galactic diffuse models challenging.

\begin{table}[t!]
\centering\renewcommand{\arraystretch}{1.22}
\begin{tabular}{cccc}
\hline
\makebox[1.5cm][c]{Spiral Arm} & \makebox[1.5cm][c]{$\beta_{i}$} & \makebox[1.5cm][c]{$a_{i}$} &
\makebox[1.5cm][c]{$w_{i}$} \\
\hline
Sagittarius-Carina & 0.242 & 0.246 & 0.178
 \\
Scutum-Crux & 0.279 & 0.608 &  0.280 \\
Norma-Cygnus & 0.249 & 0.449 & 0.357 \\
Perseus & 0.240 & 0.378 & 0.185 \\
\hline
\end{tabular}
\caption[]{Parameters of the Galactic spiral arms in Eq.~(\ref{eq:spiral}) based on Ref.~\cite{Steiman-Cameron:2010iuq} with names following the common naming conventions. The relative weights $w_i$ of the arms are normalized so that $\sum_iw_i=1$.}\label{tab1}
\end{table}

\begin{table}[t!]
\centering\renewcommand{\arraystretch}{1.22}
\begin{tabular}{cccc}
\multicolumn{4}{c}{\begin{minipage}[c][0.8cm][c]{0.8\linewidth}SNR Catalog of Ref.~\cite{IceCube2023galactic}\end{minipage}}\\
\begin{minipage}[c][0.5cm][c]{2cm}Source name\end{minipage} & \makebox[1.7cm][c]{$\ell$ [$^\circ$]} & \makebox[1.7cm][c]{$D^{\rm min}$ [kpc]} &
\makebox[1.7cm][c]{$D^{\rm max}$ [kpc]} \\
\hline
HESS J1912+101    & $44.5$ & 3.4 & 4.8 \\
Gamma Cygni       & $78.2$ & 0.49 &1.96  \\
Cassiopeia A      & $111.7$ & 3.3 & 3.7 \\
Vela Junior       & $266.2$ & 0.5 & 1 \\
RCW 86            & $315.1$ & 2.1 & 3.2 \\
HESS J1457-593    & $318.2$ & 2.3 & 3.17 \\
SNR G323.7-01.0   & $323.6$ & 3.5 & 6 \\
HESS J1614-518    & $331.5$ & - & - \\
RX J1713.7-3946   & $347.3$ & 0.9 & 1.1 \\
CTB 37A           & $348.4$ & 6.3 & 12.5 \\
HESS J1731-347    & $353.5$ & 2.4 & 6.1 \\
HESS J1745-303    & $358.8$ & 3.13 &3.85  \\
\hline
\multicolumn{4}{c}{\begin{minipage}[c][0.8cm][c]{0.8\linewidth}PWN Catalog of Ref.~\cite{IceCube2023galactic}\end{minipage}}\\
\begin{minipage}[c][0.5cm][c]{2cm}Source name\end{minipage} & \makebox[1.5cm][c]{$\ell$ [$^\circ$]} & \makebox[1.5cm][c]{$D^{\rm min}$ [kpc]} &
\makebox[1.5cm][c]{$D^{\rm max}$ [kpc]} \\
\hline
HESS J1813-178 & $12.8$ & 2      & 8 \\
HESS J1825-137 & $18.0$ & 2.9    & 3.3 \\
HESS J1837-069 & $25.2$ & 3.6    & 4 \\
Crab Nebula    & $184.5$ & 1.7    & 2.1 \\
Vela X         & $263.3$ & 0.25   & 0.3 \\
HESS J1026-582 & $285.1$ & 2.5    & 10 \\
HESS J1303-631 & $304.2$ & 0.75   & 3 \\
Kookaburra     & $313.3$ & 2.8    & 11.2 \\
MSH 15-52      & $320.3$ & 3.8    & 6.6 \\
HESS J1616-508 & $332.3$ & 2.7    & 3.3 \\
HESS J1632-478 & $336.3$ & 1.5    & 6 \\
HESS J1708-443 & $343.1$ & 1      & 4 \\
\hline
\multicolumn{4}{c}{\begin{minipage}[c][0.8cm][c]{0.8\linewidth}List of YMSCs shown in Fig.~\ref{fig:DP_MW}\end{minipage}}\\
\begin{minipage}[c][0.5cm][c]{2cm}Source name\end{minipage} & \makebox[1.5cm][c]{$\ell$ [$^\circ$]} & \makebox[1.5cm][c]{$D^{\rm min}$ [kpc]} &
\makebox[1.5cm][c]{$D^{\rm max}$ [kpc]} \\
\hline
Arches                  & $0.1$    & 7      & 9  \\
Quintuplet              & $0.2$    & 7      & 9  \\
RSGC01                  & $25.3$   & 5.05   & 7.73  \\
RSGC02                  & $26.2$   & 5.05   & 7.74  \\
Cygnus OB2              & $80.2$   & 1.24   & 1.94  \\
h Per (NGC 869)         & $134.6$  & 1.84   & 2.44  \\
$\chi$ Per (NGC 884)    & $135.0$  & 1.84   & 2.44  \\
NGC 3603                & $283.7$  & 6.6    & 8.6  \\
Westerlund 2            & $284.3$  & 3.84   & 4.48  \\
Trumpler 14             & $287.4$  & 2.6    & 3.4  \\
Westerlund 1            & $339.6$  & 3.54   & 3.72   \\
DBS2003 179             & $347.6$  & 6.9    & 9.1  \\
\hline
\end{tabular}
\caption[]{List of SNRs and PWNe studied in Ref.~\cite{IceCube2023galactic} in ascending Galactic longitude $\ell$. The distance range is taken from Ref.~\cite{Ferrand:2012jh}. For sources where only an approximate distance $D$ is provided, we follow the approach of Ref.~\cite{HESS:2018pbp} with $D^{\rm min} = D/2$ and $D^{\rm max} = 2D$. Also listed are nearby YMSCs~\cite{Kharchenko:2005fe,Pfalzner:2009vn,Morales,Alvarez:2013ig} indicated in Fig.~\ref{fig:DP_MW}.}\label{tab2}
\end{table}

\section{Galactic Source Distribution} \label{app:distribution}

Our discussion of QD neutrino emission focused on azimuthally symmetric Galactic source distributions. Here, we also consider source distributions following the Galactic arm structure. As our benchmark model we choose the 4-arm spiral model provided in Ref.~\cite{Steiman-Cameron:2010iuq} with the parametrization:
\begin{equation}\label{eq:spiral}
\rho(r,\phi,z) \equiv\rho(r)\sum_i w_i \frac{e^{\kappa\cos{(\phi-\phi_i(r))}}}{I_0(\kappa)}e^{- \frac{z^2}{2\sigma_z^2}}\,,
\end{equation}
where $\phi_i(r) \equiv \ln(r/a_i)/\beta_i$, $R = 2.9$~kpc, $\sigma_z = 0.07$~kpc, $\kappa = 29.1$, and the parameters $a_i$, $\beta_i$ and $w_i$ are given in Table~\ref{tab1}. This model uses a Galactocentric coordinate system where the Solar System is located at $x_\odot=0$ and $y_\odot=8.5$~kpc for $x\equiv r\cos(\phi)$ and $y \equiv r\sin(\phi)$. (Note that the visualizations of the 4-arm spiral in Figs.~\ref{fig:DP_MW} and \ref{fig:4arm} uses a rotated coordinate system so that the Solar System appears at $x_\odot=0$ and $y_\odot=-8.5$~kpc.) The parametrization (\ref{eq:spiral}) deviates from that of Ref.~\cite{Steiman-Cameron:2010iuq} that uses a Gaussian distribution in azimuthal angle $\phi$. Our version (\ref{eq:spiral}) makes use of normalized von Mises distributions that are approximately Gaussian in the limit of small widths $\kappa^{-1/2}$. Our normalization of Eq.~(\ref{eq:spiral}) ensures that:
\begin{equation}
\frac{1}{2\pi}\int_0^{2\pi}{\rm d}\phi\rho(r,\phi,0) = \rho(r)\,,
\end{equation}
where $\rho(r)$ is the radial distribution of Eq.~(\ref{eq:galactic_density}). 
Again, varying $\alpha$ and $\beta$ in Eq.~(\ref{eq:galactic_density}) within $[0,2]$ and $[1,4]$, respectively, we find that the Galactic form factor remains within $1.3\lesssim\xi_{\rm gal}\lesssim2.9$, demonstrating the robustness of our results.

\section{Galactic Sources} \label{app:sources}

Table~\ref{tab2} lists the catalogs of SNRs and PWNe that were analyzed in terms of their combined neutrino emission in Ref.~\cite{IceCube2023galactic}. The table shows the sources ordered by increasing Galactic longitude $\ell$ and with their estimated distance $D_{\rm min}<D<D_{\rm max}$. For sources with no available distance uncertainties, we estimate the uncertainty from the most probable source location $D$ as $D^{\rm min} = D/2$ and $D^{\rm max} = 2D$, following the approach of Ref.~\cite{HESS:2018pbp}. For completion, the bottom rows of Tab.~\ref{tab2} also list the nearby YMSCs~\cite{Pfalzner:2009vn,Morales, Kharchenko:2005fe,Alvarez:2013ig} shown in Fig.~\ref{fig:DP_MW}.

\begin{figure}[t!]
\centering
\includegraphics[width=\linewidth]{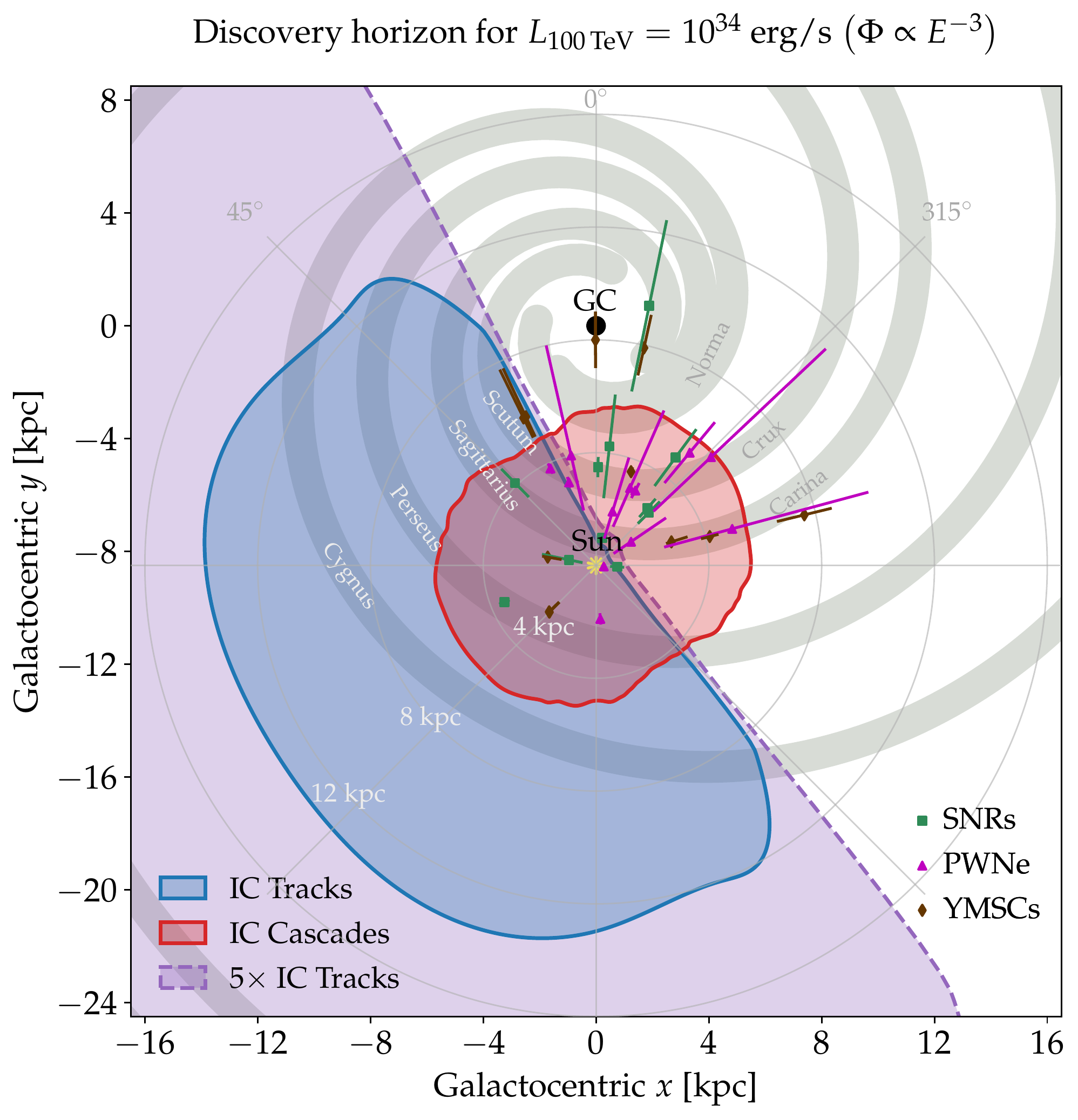}
\caption[]{Same as Fig.~\ref{fig:DP_MW} but now showing the discovery horizon of point-like sources for soft $E^{-3}$ neutrino emission. IceCube-Gen2 is here approximated as a detector with five times IceCube's DP for an $E^{-3}$ spectrum using tracks (``5$\times$IC Tracks'').}\label{fig:DP_MW_E3}
\end{figure}

Assuming a flat prior on distance, $f(D) = (D_i^{\rm max}- D_i^{\rm min})^{-1}$, we can estimate the individual source flux as:
\begin{multline}
E^2\Phi = \int\limits^{D_i^{\rm max}}_{D_i^{\rm min}}dD f(D)\frac{L_{100\,{\rm TeV}}}{4\pi D^2} = \frac{L_{100\,{\rm TeV}}}{4\pi D_i^{\rm min}D_i^{\rm max} }\,.
\end{multline}
The upper limits on the combined emission $\Phi^{90\% {\rm UL}}$ derived in IceCube's stacking searches for SNRs and PWNe imply an upper limit of the source luminosity of:
\begin{equation}\label{eq:SNRstacking}
    L_{100\,{\rm TeV}} < {E^2\Phi^{90\% {\rm UL}}_{\rm{stack}}}\left[\sum_{i=1}^{N_{\rm cat}}\frac{1}{4\pi D^{\rm min}_i D^{\rm max}_i}\right]^{-1}\,.
\end{equation}
Note that the location of the SNR candidate HESS J1614-518 (see Tab.~\ref{tab2}) is not determined. We exclude this source in the sum of Eq.~(\ref{eq:SNRstacking}), which provides a conservative upper limit on $L_{100\,{\rm TeV}}$ for SNRs.

\section{Discovery Horizon for $E^{-3}$ Spectra}

We considered in this article the emission of Galactic neutrino sources following $E^{-2}$ spectra. For completion, Fig.~\ref{fig:DP_MW_E3} presents also the discovery horizon of Eq.~(\ref{eq:Dmax}) for Galactic sources following an $E^{-3}$ spectrum. As in Fig.~\ref{fig:DP_MW}, the discovery horizon is shown for the IceCube DP for point-like neutrino source searches for track-like events \cite{IceCube:2019cia} (``IC Tracks''; blue contour) and cascade events \cite{IceCube2023galactic} (``IC Cascades''; red contour), as well as 5 times the IceCube tracks DP used here as a proxy for the DP of the proposed IceCube-Gen2 facility~\cite{IceCube-Gen2:2020qha} (``5$\times$IC Tracks''; magenta contour), assuming a monochromatic neutrino luminosity $L_{100\, \mathrm{ TeV}} = 10^{34} ~\mathrm{erg/s}$. Unfortunately, we cannot provide the DP of $E^{-3}$ spectrum for KM3NeT. 

Due to IceCube's location at the South Pole and the large background from atmospheric muons above the detector, the discovery horizon for track-like events is drastically reduced for declinations $\delta \gtrsim 5^\circ$ corresponding to zenith angles $\theta \lesssim 95^\circ$. On the other hand, cascade-like events have a more uniform coverage in declination since these are less effected by atmospheric muons.

\end{document}